\documentclass{article}

\usepackage[nonatbib,final]{neurips_2022_ml4ps}

\usepackage[utf8]{inputenc} % allow utf-8 input
\usepackage[T1]{fontenc}    % use 8-bit T1 fonts
\usepackage[hidelinks]{hyperref}       % hyperlinks
\usepackage{url}            % simple URL typesetting
\usepackage{booktabs}       % professional-quality tables
\usepackage{amsfonts}       % blackboard math symbols
\usepackage{nicefrac}       % compact symbols for 1/2, etc.
\usepackage{microtype}      % microtypography
\usepackage{xcolor}         % colors
\usepackage{float}
\usepackage{bbm}
\usepackage{tabularx}
\usepackage{lineno}
\usepackage{changepage}
\usepackage{adjustbox}
\usepackage{amsmath,amsthm,amssymb,amsfonts, fancyhdr, color, comment, graphicx, environ}
\usepackage{todonotes}

\usepackage{amsmath}

\DeclareMathOperator*{\argmin}{arg\,min}
\DeclareMathOperator{\logit}{logit}

\usepackage{setspace}
\title{Scalable Bayesian Inference for Finding Strong Gravitational Lenses}

\author{%
  Yash Patel \\
  Department of Statistics\\
  University of Michigan\\
  \texttt{yppatel@umich.edu} \\
   \And
  Jeffrey Regier \\
  Department of Statistics\\
  University of Michigan\\
  \texttt{regier@umich.edu} \\
}



\begin{document}

\maketitle

\begin{abstract}
  Finding strong gravitational lenses in astronomical images allows us to assess cosmological theories and
  understand the large-scale structure of the universe. Previous works on lens detection % detecting strong lenses 
  do not quantify uncertainties in lens parameter estimates or 
  scale to modern surveys. We present a fully amortized Bayesian procedure for lens detection 
  % performing detection of gravitational lenses 
  that overcomes these limitations. Unlike traditional variational inference, in which training minimizes the reverse Kullback-Leibler (KL) divergence, our method is trained with an expected forward KL
  divergence. Using synthetic GalSim images and real Sloan Digital Sky Survey (SDSS) images, we demonstrate that amortized inference trained with the forward KL produces well-calibrated uncertainties in both lens detection and parameter estimation.
\end{abstract}

\section{Introduction}
Strong gravitational lensing events
% , where light is visibly deflected from a source galaxy, 
are widely used to validate and parameterize the $\Lambda$CDM model, the current concordance model in cosmology \cite{gilman2021strong, hezaveh2016detection, vegetti2010detection, vegetti2009statistics, hogg1994gravitational}.
% The concordance model in cosmology is the $\Lambda$CDM model
% %, representing our best conception of the universe at cosmological scales 
% \cite{blanchard2022lambda, bull2016beyond}. For this reason, there is great interest in its accurate parameterization. Such parameterization is tested using data from strong lensing events, where light is visibly deflected from a source galaxy \cite{gilman2021strong, hezaveh2016detection, vegetti2010detection, vegetti2009statistics, hogg1994gravitational}. 
% Dark matter, however, cannot be observed and hence the properties of interest must be inferred through indirect means, specifically through the observed effects on light by such dark matter in the form of lensing events. 
% Strong lensing, where light is visibly deflected from a source galaxy, is one such detectable event, from which dark matter distributions can be inferred \cite{vegetti2009statistics, hogg1994gravitational}.
Despite extensive study, it remains challenging to efficiently detect strong lenses and to accurately estimate their characteristics.
% computationally tractable manner. 
Because researchers anticipate that the upcoming Legacy Survey of Space and Time (LSST) will image roughly $10^5$ lenses \cite{liu2020testing}, such efficient detection is of interest. 

Non-generative deep learning detectors are computationally efficient \cite{hezaveh2017fast, brehmer2019mining, alexander2020deep}, but they sacrifice the accuracy and uncertainty quantification provided by fully generative models. Additionally, they do not cope well with blending, instances where \textit{multiple} galaxies overlap visually. 
% in the imaged data. 
Handling blending is paramount, as it is anticipated that 62\% of imaged galaxies in LSST will be blended \cite{sanchez2021effects}. Standalone deblenders have been developed \cite{melchior2018scarlet, lanusse2021deep}, but they lack probabilistic interpretability.
Bayesian methods have been shown to address these deficiencies yet remain too computationally demanding to be run on large datasets. A recent work 
\cite{gu2022giga} employs a bespoke sampling process 
to improve upon Hamiltonian Monte Carlo 
yet still requires 105 seconds on four Nvidia A100 GPUs for a single lens. Further, previous works that used variational inference were trained with the reverse KL divergence, which is known to produce underdispersed posterior approximations \cite{pml2Book}. A separate line of work has studied lens substructure \cite{chianese2020differentiable,mishra2022strong}; however, these works presume the identification of a lens. 
% starting by identifying regions of high posterior density for HMC initialization, followed by computing the posterior covariance matrix $\Sigma$, and finally using $\Sigma^{-1}$ as a preconditioner to generate HMC samples \cite{gu2022giga}. 
% Despite significant implementation optimization, however, this method remains computationally intractable, requiring 105 seconds on four NVIDIA A100 GPUs for a single lens. Further, these aforementioned works are all trained with the reverse KL divergence, which is known to produce underdispersed posterior approximations \cite{pml2Book}. 

We propose to detect strong lensing while fitting a generative model for deblending with an inference procedure that is scalable to modern astronomical surveys. Our method is amortized, supports calibrated uncertainty quantification, and is available at \url{https://github.com/prob-ml/bliss}.

\section{Statistical Model}
Astronomical images record radiation originating from light sources, such as stars and galaxies. Catalogs contain properties of these sources, such as their locations and fluxes. It is of interest to infer the posterior distributions for these properties in addition to those of strong gravitational lenses. 

We propose the following generative model for this task, which extends the BLISS model \cite{liu2021variational,hansen2022scalable}.
First, draw the number of imaged light sources from a Poisson process, $S\sim \text{Poisson}(\mu\eta),$
% \begin{equation}
%     S\sim \text{Poisson}(\mu\eta),
% \end{equation} 
with $\mu$ denoting the average density of light sources per square degree of the image and $\eta$ denoting the number of square degrees. 
% Lensing is taken to be a property of a galaxy, meaning there are no ``additional'' galaxies needed to be to drawn from this spatial Poisson process to allow for its manifestation. 
% This does have the limitation of restricting lensing of a given lens galaxy to just a single other galaxy, which is accurate for sufficiently separated galaxies, but may be of additional interest to model if such events are found to be common in emerging astronomical surveys. 
Then, for each source $s = 1,...,S$, the location and type of the source are

% \begin{align}
% c_s \mid S\sim\text{Unif}([0, H]\times[0, W])\\
% a_s \mid S \sim \text{Bernoulli}(\rho_s),
% \end{align}

% \noindent\begin{minipage}{.5\linewidth}
% \begin{equation}
%   c_s \mid S\sim\text{Unif}([0, H]\times[0, W])
% \end{equation}
% \end{minipage}%
% \begin{minipage}{.5\linewidth}
% \begin{equation}
%   a_s \mid S \sim \text{Bernoulli}(\rho_s),
% \end{equation}
% \end{minipage}

\begin{equation}
  u_s \mid S\sim\text{Unif}([0, H]\times[0, W])
  \qquad\text{and}\qquad
  a_s \mid S \sim \text{Bernoulli}(\rho_s),
\end{equation}

where $\rho_s$ is the proportion of sources that are stars, and $1-\rho_s$ is the proportion that are galaxies. We use the star and galaxy flux models presented in \cite{hansen2022scalable}, namely TruncatedPareto($f_{\min}, 0.5$) for stars and a bulge-and-disk model $\mathcal{G}$ for galaxies, parameterized by $g_s$. The full specification of $g_s$ is given in Table~\ref{tab:params}.
% For notational clarity, throughout the remainder of this paper, we do not explicitly notate the dependence of latent variables on $s$, although \textbf{all} latent variables are modeled \textit{per source} (i.e. separately per $s$). 
% Taking $f_{\min}$ to be the minimum flux for a star that can be detected \cite{hansen2022scalable}, the flux is described by a TruncatedPareto($f_{\min}, \alpha$), with $\alpha=0.5$. If, on the other hand, the light source is galaxy, we represent its spatial properties with a bulge-and-disk model $\mathcal{G}$, characterized by $g_s := (f_T, d_p, \beta, d_q, b_q, a_d, a_b)_s$ (parameters defined in Table~\ref{tab:params}) \cite{hansen2022scalable}. 
Additionally, if a source is a galaxy, whether it is lensed is indicated by

% \begin{equation}
% f_{1,s} \mid (S, a_s = 1) \sim \text{TruncatedPareto}(f_{\min}, \alpha),
% \end{equation}

% With $\alpha=0.5$. 

% This is for the first band. For other bands, we use:

% \begin{equation}
% c_{s,b} \mid S, a_s = 1 \sim \mathcal{N}(\mu_c, \sigma^2_c^2)
% \end{equation}

% as (where $d,b$ are respectively the disk and bulge parameters)
% , where $f_T$ is the total flux of the galaxy, $d_p$ the proportion of the flux taken by the disk component, $\beta$ the angular offset, $d_q$ the minor-to-major axis ratio for the disk, $b_q$ the minor-to-major axis ratio for the bulge, $a_d$ the major axis for the disk, and $a_b$ the major axis for the bulge. 
% For the full prior on these parameters, refer to Table~\ref{tab:params}.

% We place the following priors on these parameters:

% \begin{align}
% f_{T,s} \sim   \text{Pareto}(f_{\min}, \alpha_f)\\
% d_{p,s} \sim   \mathcal{U}[0,1] \\
% \beta_{s} \sim \mathcal{U}[0,2\pi] \\
% d_{q,s} \sim   \mathcal{U}[0,1]\\
% b_{q,s} \sim   \mathcal{U}[0,1]\\
% a_{d,s} \sim   \text{Gamma}(\alpha_{d}, \beta_{d}) \\
% a_{b,s} \sim   \text{Gamma}(\alpha_{b}, \beta_{b})
% \end{align}

% Therefore, the flux model for a galaxy is modeled as:

% \begin{equation}
%     \mathcal{G}(g) := 
%     \sum_{\tau\in\{d,b\}} \text{Sérsic}(I_0=f_T \tau_p, n=n_\tau, q=\tau_q, hlr=a_\tau\sqrt{\tau_q}).
% \end{equation}

\begin{equation}
\gamma_{s} \mid (S, a_s = 0) \sim \text{Bernoulli}(\rho_{\ell}),
\end{equation}

where $\rho_\ell$ is the proportion of galaxies that are lensed. We assume that all lensing events require a pair of galaxies $(s, s')$, with $s$ acting as a lens and $s'$ being lensed. The galaxy $s'$ is initially rendered unlensed (with $g_{s'}$), followed by a resampling operation on a grid warped by the singular isothermal ellipsoid (SIE) lensing potential, parameterized by $r_{\ell} := (\theta_E, q_1, q_2, \theta_x, \theta_y)$, whose values are determined by interactions between $s$ and $s'$ \cite{narayan1996lectures}.
% Henceforth, we drop the explicit dependence of $r_{\ell,s,s'}$ on $s$ and $s'$ in our notation. 
Denoting the grid distortion as $\mathcal{D}_{r_{\ell}}$, a lens pair is rendered as

% For the full prior on these parameters, refer to Table~\ref{tab:params}. Note that the bulge-and-disk parameters of the lensed galaxy are \textit{separate} from those of the lens galaxy; a total of 12 additional parameters are necessary for rendering a galaxy with lensing present, with 7 for the \textit{lensed} galaxy spatial characteristics ($g_{\ell_s}$) and 5 for its SIE lens parameters ($r_{\ell,s}$). 

\begin{equation}
    f_{0,s,s'} \mid (S, a_s = 0, a_{s'} = 0, \gamma_{s'} = 1, g_{s}, g_{s'}, r_{\ell}) = \mathcal{G}(g_{s}) + \mathcal{D}_{r_{\ell}}\left(\mathcal{G}(g_{s'})\right).
    % f_{0,s,s'} \mid (S, a_s = 0, a_{s'} = 0, \gamma_{s'} = 1, g_{s}, g_{s'}, r_{\ell}) = \underbrace{\mathcal{G}(g_{s})}_\text{Lens galaxy} + \underbrace{\mathcal{D}_{r_{\ell}}\left(\mathcal{G}(g_{s'})\right) }_\text{Lensed galaxy}.
\end{equation}

Denote the background photon contribution as $\zeta_n$, the contribution from a source $s$ to pixel $n$ as $\lambda_{n,s}$, and the complete set of latent variables as $z$. See Table~\ref{tab:params} for descriptions and priors of such parameters. Then, the number of photon arrivals observed at pixel $n$ is 

\begin{equation}
x_n \mid z\sim\text{Poisson}\left(\zeta_n + \sum_{s=1}^S \lambda_{n,s}\right).
\end{equation}

\begin{table}%[H]
\centering
\begin{adjustbox}{tabular=clll,center}
 \textbf{Name} & \textbf{Generative} & \textbf{Variational} & \textbf{Description} \\
 \hline
 \hline
  $S$             & Poisson$(\mu \eta)$ & Categorical & Number of sources \\
 $u$       & $\mathcal{U}([0, H] \times [0, W])$ & $\log(u)\sim\mathcal{N}(\mu_{u}, \sigma^2_{u})$ & Location of source \\
 $a$           & Bernoulli$(\rho_s)$ & $\text{Bernoulli}(\mu_{a_s})$ & Type of source \\
 \hline
 $f_{1}$      & Pareto$(f_{\min}, \alpha)$ & $\log(f_1)\sim\mathcal{N}(\mu_{f_1}, \sigma^2_{f_1})$ & Star flux \\
 \hline
 $f_{T}$     & $\text{Pareto}(f_{\min}, \alpha_f)$ & $\log(f_T)\sim\mathcal{N}(\mu_{f_T}, \sigma^2_{f_T})$ & Total galactic flux \\
 $d_{p}$      & $\mathcal{U}[0,1]$ & $\logit(d_p)\sim\mathcal{N}(\mu_{d_p}, \sigma^2_{d_p})$ & Disk flux proportion \\
 $\beta$      & $\mathcal{U}[0,2\pi]$ & $\logit\left(\frac{\beta} {2\pi}\right)\sim\mathcal{N}(\mu_{\beta}, \sigma^2_{\beta})$ & Galaxy ellipse angular offset \\
 $d_{q}$      & $\mathcal{U}[0,1]$ & $\logit(d_q)\sim\mathcal{N}(\mu_{d_q}, \sigma^2_{d_q})$ & Disk minor-to-major axis ratio \\
 $b_{q}$      & $\mathcal{U}[0,1]$ & $\logit(b_q)\sim\mathcal{N}(\mu_{b_q}, \sigma^2_{b_q})$ & Bulge minor-to-major axis ratio \\
 $a_{d}$      & $\text{Gamma}(\alpha_{d}, \beta_{d})$ & $\log(a_d)\sim\mathcal{N}(\mu_{a_d}, \sigma^2_{a_d})$ & Major axis for the disk \\
 $a_{b}$      & $\text{Gamma}(\alpha_{b}, \beta_{b})$ & $\log(a_b)\sim\mathcal{N}(\mu_{a_b}, \sigma^2_{a_b})$ & Major axis for the bulge \\
 $f_{0}$      & Composite Sérsic & N/A & Galaxy flux \\
 \hline
 $\gamma$ & $\text{Bernoulli}(\rho_\ell)$ & $\text{Bernoulli}(\mu_{\gamma})$ & Indicator of lensing \\
 $\theta_{E}$ & $\mathcal{U}[\theta_{E,\min},\theta_{E,\max}]$ & $\log(\theta_{E})\sim\mathcal{N}(\mu_{\theta_{E}}, \sigma^2_{\theta_{E}})$ & Einstein radius \\
 $\theta_{x}$ & $\mathcal{N}(0, 1)$ & $\theta_{x}\sim\mathcal{N}(\mu_{\theta_{x}}, \sigma^2_{\theta_{x}})$ & Lens center $x$ \\
 $\theta_{y}$ & $\mathcal{N}(0, 1)$ & $\theta_{y}\sim\mathcal{N}(\mu_{\theta_{y}}, \sigma^2_{\theta_{y}})$ & Lens center $y$ \\
 $q_{\ell}$ & $\mathcal{U}[0,1]$ & N/A & Lens minor-to-major axis ratio \\
 $\beta_{\ell}$ & $\mathcal{U}[-\pi/4,\pi/4]$ & N/A & Lens angular offset\\
 $e_{1}$ & $\frac{1-q_{\ell}}{1+q_{\ell}}\cos(\beta_{\ell})$ & $\logit\left(\frac{e_1 + 1}{2}\right)\sim\mathcal{N}(\mu_{e_1}, \sigma^2_{e_1})$ & Lens ellipticity (factor 1) \\
 $e_{2}$ & $\frac{1-q_{\ell}}{1+q_{\ell}}\sin(\beta_{\ell})$ & $\logit\left(\frac{e_2 + 1}{2}\right)\sim\mathcal{N}(\mu_{e_2}, \sigma^2_{e_2})$ & Lens ellipticity (factor 2) \\
 \hline
\end{adjustbox}
\caption{\label{tab:params} Parameters for the generative model and variational distribution. The four partitions of the table respectively correspond to the detection, star, galaxy, and lens parameters.}
\end{table}

\section{Variational Inference}
% \label{inference}
We aim to minimize the expected forward KL divergence to approximate the posterior distribution using forward amortized variance inference (FAVI) \cite{ambrogioni2019forward}. We thus aim to solve

% We split the image into tiles and perform variational inference across such tiles, making the goal to find

\begin{equation}
    \argmin_{\varphi}
    \mathbb{E}_{(x,z)\sim p(z)p(x|z)}\left[\log(q_\varphi(z|x))\right].
    % =
    % \argmin_{\varphi} \mathbb{E}_{(x,z)\sim p(z)p(x|z)}\left[\sum_{t=1}^T \log(q_\varphi(z_t|x))\right].
\end{equation}

% For this, we wish to find 
% the $\argmin_{\varphi}$ of

% \begin{align}
%     \mathbb{E}_{x\sim p(x)}\left[
%     D_{\mathbb{K}\mathbb{L}}(p_\theta(z | x) || q_\varphi(z | x))
%     \right]
%     = \mathbb{E}_{(x,z)\sim p(z)p(x|z)}\left[
%     \log\left(\frac{p_\theta(x,z)}{q_\varphi(z|x)}\right)
%     \right].
% \end{align}

% Notice we can reframe this objective by eliminating terms that are not of relevance for optimization as the following:

% \begin{equation}
%     \argmin_{\varphi} -\mathbb{E}_{x\sim p(x)}\left[
%     D_{\mathbb{K}\mathbb{L}}(p_\theta(z | x) || q_\varphi(z | x))
%     \right]
%     = \argmin_{\varphi}
%     \mathbb{E}_{(x,z)\sim p(z)p(x|z)}\left[\log(q_\varphi(z|x))\right]
% \end{equation}

% For additional computational efficiency, we split the image into tiles and perform variational inference across such tiles. The final loss employed, therefore, is
    
% \begin{equation}
%     \mathbb{E}_{(x,z)\sim p(z)p(x|z)}\left[\sum_{t=1}^T \log(q(z_t|x))\right].
% \end{equation}

Because we employ the FAVI loss, we are not restricted to reparameterizable distributions. We thus use the following variational distribution:

\begin{equation}
\label{eqn:var_dist}
    q_\varphi(z | x)
    = q(S) 
    \prod_{s=1}^S 
    q(\ell_{s} | S)
    q(a_{s} | S)
    q(g_{s} | S, a_{s})
    q(\gamma_{s} | S, a_{s})
    q(r_{\ell,s} | S, a_{s}, \gamma_{s}).
\end{equation}

Table~\ref{tab:params} gives the distributional form of each factor. Each factor was approximated using a separate ``encoder'' neural network, one for each of the following tasks: source count estimation, source classification, galaxy parameter estimation, lens classification, and lens parameter estimation.

\section{Results}
All encoders were implemented in PyTorch \cite{paszke2019pytorch} with standard CNN architectures and employed the tiling decomposition described in \cite{hansen2022scalable}.
% ReLU, dropout, and batch normalization layers
%\footnote{Full details of architectural and training hyperparameter choices are available in the code.}.
Each was trained separately on synthetic images from the generative model, with galaxies rendered using GalSim~\cite{rowe2015galsim}. Optimization was done using Adam \cite{kingma2014adam}. Training these models required five hours using eight Nvidia RTX 2080 Ti GPUs. This is a one-time cost: inference can be run on an arbitrary number of images thereafter without additional training. We used both synthetic data and images from SDSS for validation. 
% Standard CNN architectures were employed for each of the encoders, with ReLU non-linearities, dropout, and batch normalization layers. The complete set of hyperparameters are available in the code.

\subsection{Synthetic Images}
\begin{figure}[H]
\centering
\includegraphics[scale=0.24]{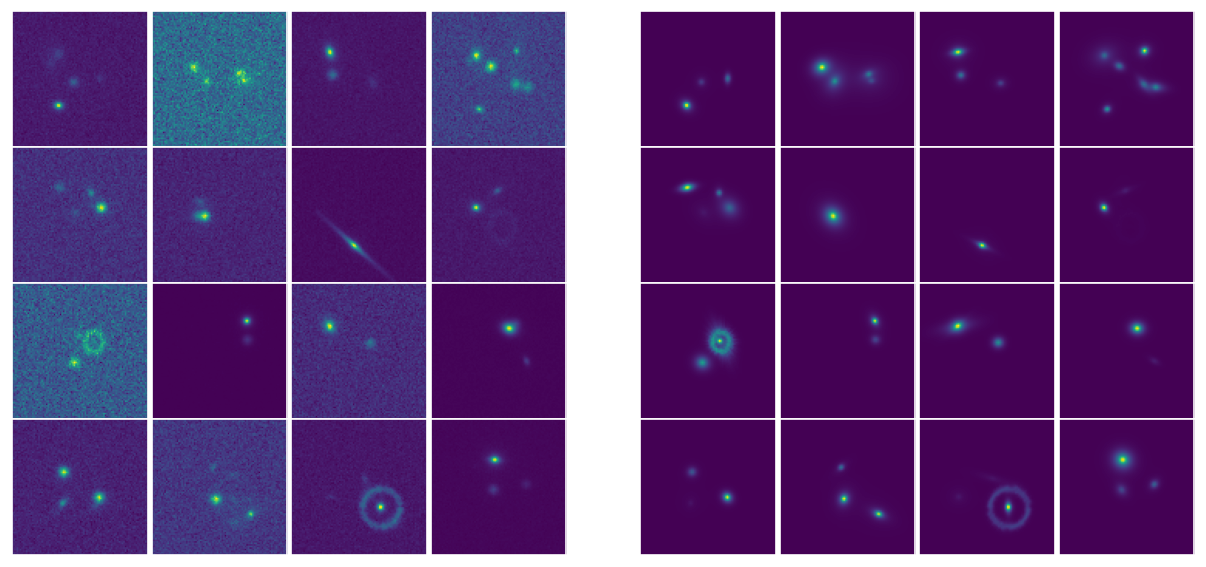}
\caption{\label{fig:sims} Synthetic images from our generative model. Each is normalized against the brightest object in the image. The left panel shows the original synthetic images and the right our reconstructions. Checking the similarity of the reconstructed images serves as an initial qualitative posterior check.}
\end{figure}

\begin{figure}[tbp]
  \centering
  \begin{minipage}[b]{1.0\linewidth}
    \centering
    \includegraphics[width=0.3\textwidth]{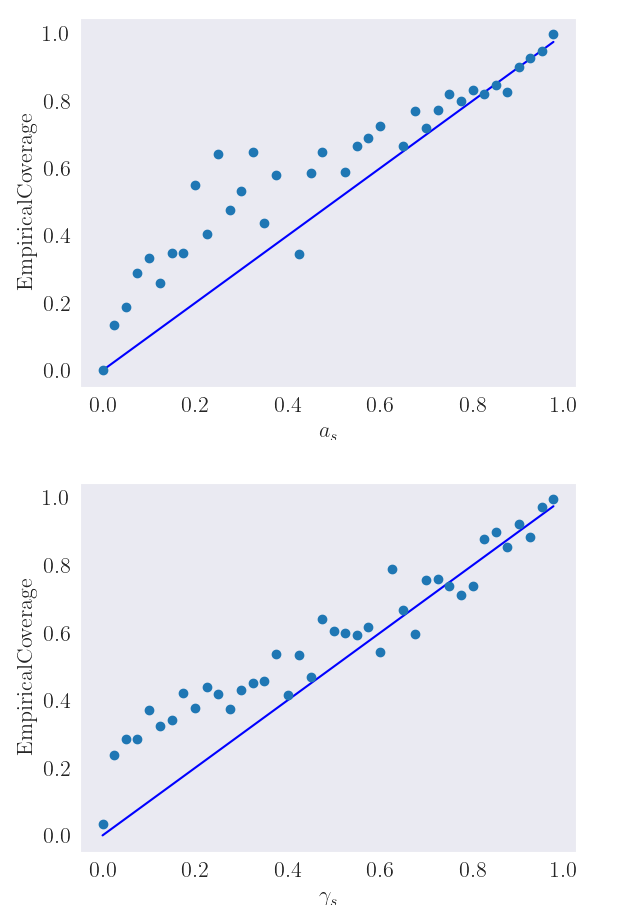}
    \qquad
    \begin{tabular}[b]{cc|cc}
     \textbf{Name} & \textbf{Coverage} & \textbf{Name} & \textbf{Coverage} \\
     \hline
     \hline
     $f_{T}$   & 94.66 \% & $f_{T,\ell}$   & 91.22\% \\
     $d_{p}$   & 89.62 \% & $d_{p,\ell}$   & 87.79\% \\
     $\beta_s$ & 90.88 \% & $\beta_{s,\ell}$ & 89.87\% \\
     $d_{q}$   & 88.11 \% & $d_{q,\ell}$   & 89.05\% \\
     $b_{q}$   & 87.29 \% & $b_{q,\ell}$   & 89.42\% \\
     $a_{d}$   & 87.24 \% & $a_{d,\ell}$   & 89.24\% \\
     $a_{b}$   & 90.20 \% & $a_{b,\ell}$   & 95.47\% \\
     $\theta_{E}$ & 94.39 \% & $e_1$   & 94.66\% \\
     $\theta_{x}$ & 92.22 \% & $e_2$   & 89.78\% \\
     $\theta_{y}$ & 91.50 \% & & \\
    \end{tabular}
  \end{minipage}
  \caption{Assessment of posterior calibration for detection and continuous parameters, respectively shown in the graph and table. The blue lines in the graphs represent the ideal calibrations.
Bayes credible intervals were constructed for 90\% coverage.}
    \label{fig:posterior_cal}
\end{figure}

% Using latent draws from our model, we obtain samples akin to those seen in Figure~\ref{fig:sims}. 
The encoders were trained on data generated through the posited forward model, as shown in Figure~\ref{fig:sims}. Post-training validation was also performed in a number of ways. In particular, Figure~\ref{fig:sims} also serves as a visual qualitative posterior check. To assess uncertainty calibration, discrete and continuous latent quantities were handled separately. Discrete quantities, namely galaxy and lens detection, were plotted with their outputted posterior probabilities against the empirical proportions. Continuous variable posterior calibration was assessed with coverage percentages for 90\% Bayes credible intervals. Results in Figure~\ref{fig:posterior_cal} reveal well-calibrated posterior distributions for detection and parameter estimates for both galaxies and lenses. Understanding specific sources of calibration imperfections is of interest; one plausible cause stems from the limited expressivity of the encoders, owing to the fact the neural networks have a finite number of layers.

% \begin{figure}[H]
% \centering
% \includegraphics[scale=0.23]{images/posteriors2.png}
% \caption{\label{fig:posterior_cal} Assessment of posterior calibration for discrete (detection) and continuous latent variables, respectively shown in the graph and table.  The blue lines in the graphs represent the ideal calibration.}
% \end{figure}

% \begin{table}[H]
% \centering
% \begin{adjustbox}{tabular=cc|cc,center}
%  \textbf{Name} & \textbf{Coverage} & \textbf{Name} & \textbf{Coverage} \\
%  \hline
%  \hline
%  $f_{T}$   & 94.66 \% & $f_{T,\ell}$   & 91.22\% \\
%  $d_{p}$   & 89.62 \% & $d_{p,\ell}$   & 87.79\% \\
%  $\beta_s$ & 90.88 \% & $\beta_{s,\ell}$ & 89.87\% \\
%  $d_{q}$   & 88.11 \% & $d_{q,\ell}$   & 89.05\% \\
%  $b_{q}$   & 87.29 \% & $b_{q,\ell}$   & 89.42\% \\
%  $a_{d}$   & 87.24 \% & $a_{d,\ell}$   & 89.24\% \\
%  $a_{b}$   & 90.20 \% & $a_{b,\ell}$   & 95.47\% \\
%  $\theta_{E}$ & 94.39 \% & $e_1$   & 94.66\% \\
%  $\theta_{x}$ & 92.22 \% & $e_2$   & 89.78\% \\
%  $\theta_{y}$ & 91.50 \% & & 
% \end{adjustbox}
% \caption{\label{tab:coverage} Empirical coverage percentages for a 90\% credible interval. Note that the galaxy parameters are listed twice (once with the $\ell$ subscript and once without) because the lens encoder must separately learn the galaxy parameters of the \textit{source} galaxy.}
% \end{table}

\subsection{Sloan Digital Sky Survey (SDSS)}
\begin{figure}[H]
\centering
\makebox[\textwidth][c]{\includegraphics[scale=0.21]{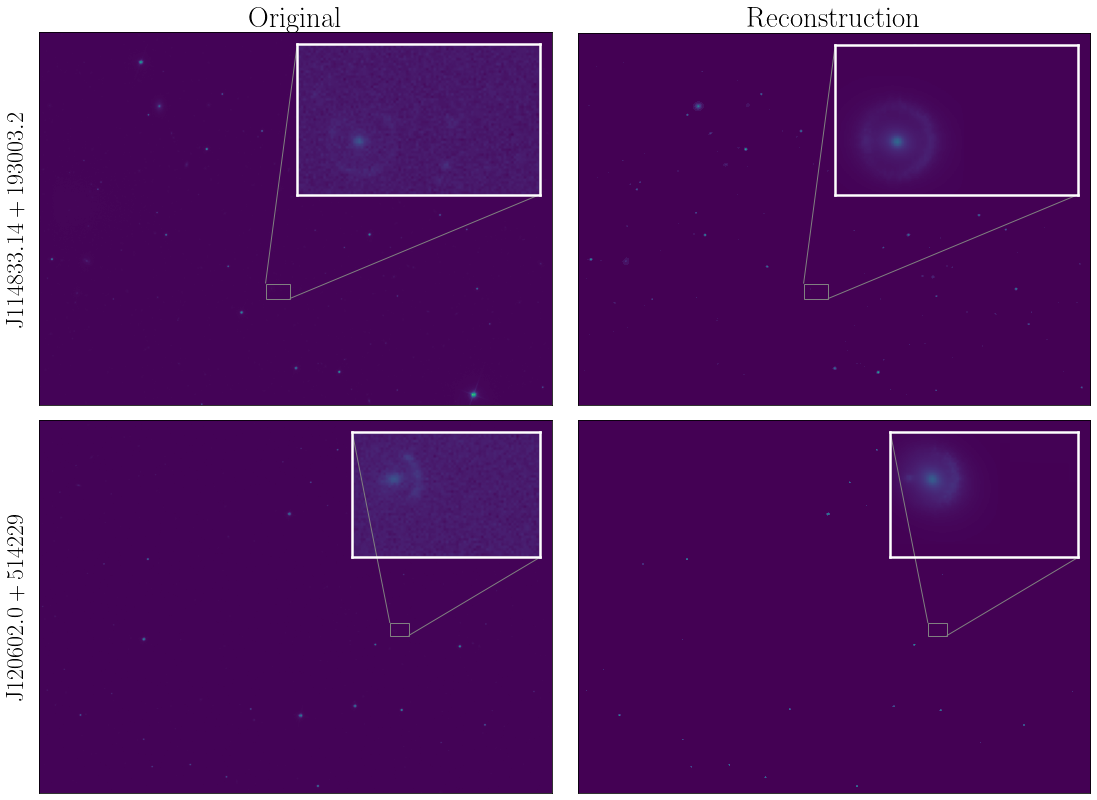}}
\caption{\label{fig:horseshoe_detection} 
% We successfully detect and reconstruct fields from SDSS. https://www.overleaf.com/project/632c6dba6b79299d48cafe37
The left column shows the original images from SDSS and the right our reconstructions. The top row is the reconstruction pair for the field containing J114833.14+193003.2 and the bottom that for J120602.0+514229. The zoom box is to highlight the subregions containing the lenses.}
\end{figure}

We additionally apply our model to two SDSS images, referencing annotations of lenses from \cite{wen2009discovery}.
% in which a number of lenses have been previously observed.
% and find both the detection and subsequent reconstruction to perform reliably. 
% Execution times for amortized inference is provided in each of the figures displayed below. 
% Annotations for these strong lenses are sourced from \cite{wen2009discovery}. 
We demonstrate successful detections in Figure~\ref{fig:horseshoe_detection}, importantly achieved \textit{without} false positives. The images were both 1489 $\times$ 2048 pixels and inference required just 25 seconds for each image. 
% Note that no additional retraining is required after the initial training.

\section{Discussion}
% We introduced a system for performing amortized inference of latent parameters for strong lenses in astronomical catalog data. 
We have shown that amortized inference performed by FAVI efficiently detects strong lenses and estimates parameters in both synthetic and real data settings while providing well-calibrated uncertainty estimates.
% to enable the analysis of astronomical data at scale. 
% With the continued expansion of such imaging technology \cite{gardner2006james}, such capabilities are of great importance for astronomers to glean insights from vast quantities of data. We expect that the use of forward KL for future work, both in this space and others where the generative model is accurate, will improve the results and calibration of generative models.
With this foundation, a number of extensions of this research are possible. One is the use of this approach to infer \textit{weak} lensing events, whose manifestations in data are quite different. For this, several non-trivial adjustments would be necessary in both the generative and inference procedures. Characterization of dark matter substructures is also of great interest and would similarly require extensions to the SIE model employed here. 
% Finally, there are extensions exploring alternate training methods from emerging work aside from the expected forward KL.
% while FAVI performed quite favorably in our experiments, some recent emerging work has studied the use of alternate training methods, with performance that exceeds even that of FAVI: future work may wish to investigate the applicability of such methods to the study of large scale astronomy data.

\section{Impact Statement}
This work builds on the use of machine learning to further astronomical and, more generally, scientific understanding, when addressing problems in which uncertainty quantification is a necessary component. Beyond direct application of the techniques presented here to the image data gathered in future astronomical surveys, particularly the recently launched JWST \cite{gardner2006james}, the broader variational inference methodology we propose could be extended to future scientific queries. 
%Given the increasingly complex nature of modern scientific inquiry, it seems likely that such coordinated efforts between human ingenuity and machine learning will play an ever-increasing role in scientific exploration going forward.

\bibliographystyle{unsrt}
\bibliography{references}

%%%%%%%%%%%%%%%%%%%%%%%%%%%%%%%%%%%%%%%%%%%%%%%%%%%%%%%%%%%%
\section*{Checklist}

%%% BEGIN INSTRUCTIONS %%%
The checklist follows the references.  Please
read the checklist guidelines carefully for information on how to answer these
questions.  For each question, change the default \answerTODO{} to \answerYes{},
\answerNo{}, or \answerNA{}.  You are strongly encouraged to include a {\bf
justification to your answer}, either by referencing the appropriate section of
your paper or providing a brief inline description.  For example:
\begin{itemize}
  \item Did you include the license to the code and datasets? \answerYes{See Section.}
  \item Did you include the license to the code and datasets? \answerNo{The code and the data are proprietary.}
  \item Did you include the license to the code and datasets? \answerNA{}
\end{itemize}
Please do not modify the questions and only use the provided macros for your
answers.  Note that the Checklist section does not count towards the page
limit.  In your paper, please delete this instructions block and only keep the
Checklist section heading above along with the questions/answers below.
%%% END INSTRUCTIONS %%%

\begin{enumerate}

\item For all authors...
\begin{enumerate}
  \item Do the main claims made in the abstract and introduction accurately reflect the paper's contributions and scope?
    \answerYes{}
  \item Did you describe the limitations of your work?
    \answerYes{}
  \item Did you discuss any potential negative societal impacts of your work?
    \answerYes{}
  \item Have you read the ethics review guidelines and ensured that your paper conforms to them?
    \answerYes{}
\end{enumerate}

\item If you are including theoretical results...
\begin{enumerate}
  \item Did you state the full set of assumptions of all theoretical results?
    \answerNA{}
        \item Did you include complete proofs of all theoretical results?
    \answerNA{}
\end{enumerate}

\item If you ran experiments...
\begin{enumerate}
  \item Did you include the code, data, and instructions needed to reproduce the main experimental results (either in the supplemental material or as a URL)?
    \answerYes{}
  \item Did you specify all the training details (e.g., data splits, hyperparameters, how they were chosen)?
    \answerYes{}
        \item Did you report error bars (e.g., with respect to the random seed after running experiments multiple times)?
    \answerNo{}
        \item Did you include the total amount of compute and the type of resources used (e.g., type of GPUs, internal cluster, or cloud provider)?
    \answerYes{}
\end{enumerate}

\item If you are using existing assets (e.g., code, data, models) or curating/releasing new assets...
\begin{enumerate}
  \item If your work uses existing assets, did you cite the creators?
    \answerYes{}
  \item Did you mention the license of the assets?
    \answerNA{}
  \item Did you include any new assets either in the supplemental material or as a URL?
    \answerNo{}
  \item Did you discuss whether and how consent was obtained from people whose data you're using/curating?
    \answerNA{}
  \item Did you discuss whether the data you are using/curating contains personally identifiable information or offensive content?
    \answerNA{}
\end{enumerate}

\item If you used crowdsourcing or conducted research with human subjects...
\begin{enumerate}
  \item Did you include the full text of instructions given to participants and screenshots, if applicable?
    \answerNA{}
  \item Did you describe any potential participant risks, with links to Institutional Review Board (IRB) approvals, if applicable?
    \answerNA{}
  \item Did you include the estimated hourly wage paid to participants and the total amount spent on participant compensation?
    \answerNA{}
\end{enumerate}

\end{enumerate}

%%%%%%%%%%%%%%%%%%%%%%%%%%%%%%%%%%%%%%%%%%%%%%%%%%%%%%%%%%%%

\end{document}